\documentclass[twocolumn]{svjour3}  

\usepackage{mathptmx}       
\usepackage{helvet}         
\usepackage{courier}        
\usepackage{makeidx}         
\usepackage{graphicx}        
\usepackage{multicol}        
\usepackage{multirow}

\usepackage{amsmath, amssymb}
\usepackage[linesnumbered,ruled,vlined]{algorithm2e}
\usepackage{xcolor}
\usepackage[english]{babel}

\usepackage[bookmarks=true, bookmarksnumbered=true, bookmarksopen=true,	
		unicode=true, colorlinks=true,
		linkcolor=blue,citecolor=blue,filecolor=blue,urlcolor=blue
		]{hyperref}
\usepackage{array}

\begin{document}

\title{A Community Role Approach To Assess Social Capitalists Visibility in the Twitter Network} 

\author{Nicolas Dugu\'e \and Vincent Labatut \and Anthony Perez}

\institute{Nicolas Dugu\'e, Anthony Perez \at Universit\'e d'Orl\'eans, LIFO EA 4022, F-45067 Orl\'eans, France\\ \email{\{anthony.perez\}\{nicolas.dugue\}@univ-orleans.fr}
\and Vincent Labatut \at Universit\'e d'Avignon, LIA EA 4128, F-84911 Avignon, France\\ \email{vincent.labatut@univ-avignon.fr}}

\maketitle

\sloppy

\begin{translation}{english}
\begin{abstract}
In the context of Twitter, social capitalists are specific users trying to increase their number of followers and interactions by any means. These users are not healthy for the service, because they are either spammers or real users flawing the notions of influence and visibility. Studying their behavior and understanding their position in Twitter is thus of important interest. It is also necessary to analyze how these methods effectively affect user visibility.
Based on a recently proposed method allowing to identify social capitalists, we tackle both points by studying how they are organized, and how their links spread across the Twitter follower-followee network. To that aim, we consider their position in the network w.r.t. its community structure. We use the concept of community role of a node, which describes its position in a network depending on its connectivity at the community level. However, the topological measures originally defined to characterize these roles consider only certain aspects of the community-related connectivity, and rely on a set of empirically fixed thresholds. We first show the limitations of these measures, before extending and generalizing them. Moreover, we use an unsupervised approach to identify the roles, in order to provide more flexibility relatively to the studied system. We then apply our method to the case of social capitalists and show they are highly visible on Twitter, due to the specific roles they hold.

\end{abstract}
\end{translation}


\section{Introduction}

The last decade has been marked by an increase in both the number of online social networking services and the number of users of such services. This observation is particularly relevant when considering Twitter, which had $200$ million accounts in April $2011$~\cite{Bosker2011} and reached $500$ millions in October $2012$~\cite{Holt2013}. 
Twitter is mostly used to share, seek and debate some information, or to let the world know about daily events~\cite{Java2007}. The amount of information shared on Twitter is considerable: there are about $1$ billion tweets posted every two and a half days~\cite{Rodgers13}. While focusing on microblogging, Twitter can be considered as a social networking service, since it includes social features. Indeed, to see the messages of other users, a Twitter user has to \emph{follow} them (i.e. make a subscription). Furthermore, a user can \emph{retweet}~\cite{Suh2010} other users' tweets, for instance when he finds them interesting and wants to share them with his followers \footnote{For a given user, a \emph{followee} (or friend in the Twitter API) is a user he subscribed to, and a \emph{follower} is a user that subscribed to him.}. Besides, users can \emph{mention} other users to draw their attention by adding \texttt{@UserName} in their message. 

Some Twitter users are trying to use these particular properties to spread efficiently some information~\cite{GVK+12}.
One of the simplest way to reach this objective is to gain as many followers as possible, since this gives a higher visibility to the user's tweets when using the network search engines~\cite{GVK+12}.
These specific users are called \emph{social capitalists}. They have been recently pointed out by Ghosh \textit{et al}.~\cite{GVK+12} in a study related to \emph{link-farming} in Twitter. They noticed in particular that users responding the most to the solicitation of spammers are in fact \emph{real}, \emph{active} users. To increase their number of followers, social capitalists use several techniques~\cite{DP14,GVK+12}, the most common one being to follow a lot of users \emph{regardless of their content}, just hoping to be followed back.

Because of this lack of interest in the content produced by the users they follow, social capitalists are not healthy for a service such as Twitter. Indeed, this behavior helps spammers gaining influence~\cite{GVK+12}, and more generally makes the task of finding relevant information harder for regular users. Identifying them and studying their behavior in Twitter are therefore two very important tasks to improve the service, since they can allow designing better search engines or functioning rules. In a recent article, Dugu\'e \& Perez~\cite{DP14} have designed a method to efficiently detect social capitalists. In order to better understand how they are organized, how really visible they are, and how their links spread across the network, we propose to characterize the position of social capitalists relatively to the community structure of the network~\cite{Dugue2014}.


In its simplest form, the community stru\-cture of a complex network can be defined as a partition of its node set, each part corresponding to a community. Community detection methods generally try to perform this partition in order to obtain densely connected groups of nodes, relatively to the rest of the network \cite{Newman2004a}. Hundreds of such algorithms have been defined in the last ten years, see \cite{Fortunato2010} for a very detailed review of the domain. The notion of community structure is particularly interesting because it allows studying the network at an intermediate level, compared to the more classic global (whole network) and local (node neighborhood) approaches.

The concept of community role is a good illustration of this characteristic. It consists in describing a node depending on the position it holds in its own community. We base our work on the Guimer\`a \& Amaral approach of the community role~\cite{Guimera2005}.  
After having applied a standard community detection method, Guimer\`a \& Amaral characterize each node according to two \textit{ad hoc} measures, each one describing a specific aspect of the community-related connectivity. 
The node role is then selected among $7$ predefined ones by comparing the two values to some empirically fixed thresholds assumed as universal. 

In this paper, we study the community roles of social capitalists within a freely-available Twitter follower-followee network provided by Cha \textit{et al}.~\cite{CHBG10}. 
In a first place, we highlight two important limitations of the community role approach described by~Guimer\`a \& Amaral~\cite{Guimera2005}. We show that the existing measures used to characterize the node's position do not take into account all aspects of the community-related external connectivity of a node. Moreover, we object the assumption of universality of the thresholds applied to the measures in order to distinguish the different node roles. The dataset we use constitutes a counter-example showing the original thresholds are not relevant for all systems.
We then explain how to tackle these limitations. We first introduce three new measures to characterize the external connectivity of a node in a more complete and detailed way. We then describe an unsupervised approach aiming at identifying the node roles without using fixed thresholds. Finally, we apply our method on the Twitter network to determine the position of social capitalists, and show they occupy specific roles in the network. 
In particular, most of them are well connected to their community, and overall a large part of them spread their links outside their community very efficiently. This gives meaningful insights regarding the actual visibility of these users. Indeed, they occupy roles leading to a high visibility in Twitter. 

We first present the concept of social capitalists in Twitter in more details (Section~\ref{sec:cap}). Next, we describe the method proposed by~Guimer\`a \& Amaral~\cite{Guimera2005} to identify the community roles of nodes (Section~\ref{subsec:original}) and provide some elements regarding its limitation (Section~\ref{subsec:participationlimits}). We then describe the solutions we propose to tackle these limitations (Section~\ref{subsec:generalizedmeasures}) and apply our method to study the roles of social capitalists in Twitter (Section~\ref{sec:results}). Finally, we discuss the works related to the notion of community role (Section~\ref{sec:related}).

\section{Social capitalists}
\label{sec:cap}

\subsection{Definition}
\label{subsec:def}
Similarly to what is observed on the Web, where site administrators perform \emph{links exchange} in order to increase their visibility, some social network users seek to maximize their number of virtual relationships. Because microblogging networks are focused on sharing information, not on developing friendship links, Twitter is particularly well-suited to observe and study this kind of behavior. 
Such users are called \emph{social capitalists} in~\cite{DP14,GVK+12} or \emph{friends infiltrators} in~\cite{Lee2010}. In the rest of paper, we call these users \emph{social capitalists}. These users exploit two relatively straightforward techniques, based on the reciprocation of the \emph{follow} link:
\begin{sloppy}
	\begin{itemize}
		\item \textbf{FMIFY} (Follow Me and I Follow You): the user ensures its potential followers that he will follow them back; 
		\item \textbf{IFYFM} (I Follow You, Follow Me): on the contrary, the user systematically follows other users, hoping to
be followed back.
	\end{itemize}
\end{sloppy}

Social capitalists are not healthy for a social networking service, since their methods to gain visibility and influence are not based on the production of relevant content and on getting a higher credibility. From this point of view, their high number of followers can be considered as undeserved, and biases all services based on the assumption that visible users produce or fetch interesting content (e.g. search or recommendation engines).

Social capitalists were introduced by Ghosh \textit{et al}~\cite{GVK+12} in a paper studying spam on Twitter. They noticed that users responding the most to the solicitations of spammers are real (i.e. neither bots nor fake accounts), active and even sometimes popular users. These users are engaged in a link exchange process such as the two described above. Using this observation, Ghosh \textit{et al.} manually constituted a list of $100000$ social capitalists --namely the most responsive ones to the solicitations of spammers. Social capitalists were also mentioned as a subset of spammers in~\cite{Lee2010,Lee2011} where authors succeed in building a robust classifier to detect spammers from regular users. In both these papers, social capitalists (called \emph{friends infiltrators} by the authors) are not specifically studied.
Furthermore, these papers neither studied the actual visibility of social capitalists nor the topological properties of their corresponding positions in the network. 

In \cite{DP14}, Dugu\'e \& Pe\-rez proposed an automatic method to detect social capitalists. In this article, our work is based on a list of social capitalists identified through this method, which is why it is presented in detail in the next section. Based on this list, we could compare the position and visibility of regular users and social capitalists through the notion of community role.


\subsection{Measures}
\label{subsec:measures}
The set of \emph{followees} and the set of \emph{followers} of a given user largely intersect when the said user applies social capitalism techniques. Based on this observation, Dugu\'e \& Perez~\cite{DP14} designed an automatic method to detect efficiently these users. It relies on three purely topological measures, i.e. it does not consider any content. The first measure, called \emph{overlap index} and introduced in~\cite{Sim43}, enables to detect potential social capitalists. The second, called \emph{ratio}, allows to determine if a given social capitalist uses the \textbf{FMIFY} or \textbf{IFYFM} principle. The third is simply the \textit{incoming degree}, and indicates if the social capitalist was successful in applying these principles. All of them are defined on the follower-followee network, which is a directed graph $G=(V,A)$ whose nodes $V$ represent users and links $A$ correspond to follower-to-followee relationships. In this network, the \textit{in-neighborhood} $N^{in}(u)=\{v \in V: (v,u) \in A\}$ of a node $u$ corresponds to the followers of the user represented by $u$, whereas his \textit{out-neighborhood} $N^{out}(u)=\{v \in V: (u,v) \in A\}$ corresponds to his followees.

The \emph{overlap index} $O(u)$ of a node $u$ is given by:
\begin{eqnarray}	
	O(u) & = & \frac{|N^{in}(u) \cap N^{out}(u)|}{\min{\{|N^{in}(u)|,|N^{out}(u)|\}}}
\label{eq:overlap}		
\end{eqnarray}
\noindent This measure is processed for all nodes, allowing to identify social capitalists. Indeed, an overlap index close to $1$ indicates the intersection of followers and followees is quite high, and so we can conclude the considered user applied either the \textbf{FMIFY} or \textbf{IFYFM} principle. On the contrary, a value close to $0$ means he is not a social capitalist. 

The ratio $r(u)$ of a node $u$ is defined as:
\begin{eqnarray}	
	r(u) = \frac{|N^{in}(u)|}{|N^{out}(u)|}
\label{eq:ratio}
\end{eqnarray}
\noindent This measure is processed for all social capitalists too, and allows to classify them more precisely. According to Dugu\'e \& Perez~\cite{DP14}, social capitalists following the \textbf{IFYFM} principle have a ratio greater than $1$ (\emph{i.e.} more followees than followers), whereas those using \textbf{FMIFY} have a ratio smaller than $1$. In both cases, the ratio is expected to be close to $1$. However, the analysis conducted in \cite{DP14} highlighted a third behavior, called \emph{passive}. Unlike other social capitalists (called \emph{active}), these passive users consider they have reached a sufficient level of influence, and therefore do not need to increase their number of followers. At this point, they stop applying the aforementioned principles, but still get more and more followers due to their high visibility. Consequently, their ratio is much smaller than $1$.

By processing the cardinalities of the incoming and outgoing neighborhoods, we obtain the in- and out-degrees, noted $d^{in}(u)=|N^{in}(u)|$ and $d^{out}(u)=|N^{out}(u)|$, respectively. The former, which corresponds to the number of followers, is used by Dugu\'e \& Perez~\cite{DP14} as a third criterion, in order to determine if a social capitalist was successful in the application of the \textbf{FMIFY} and \textbf{IFYFM} principles. They define \emph{low in-degree social capitalists} as social capitalists having between less than $10000$ followers, and \emph{high in-degree social capitalists} as those having more than $10000$ followers. The latter are efficiently gaining followers, and are considered successful, whereas the former are still less popular.

\subsection{Detection}
In~\cite{DP14}, Dugu\'e \& Perez applied their method to the 2009 data collected by Cha \textit{et al}.~\cite{CHBG10}. They empirically determined that a threshold of $0.74$ for the overlap index allows a high accuracy detection. They also added two constraints: the first is to consider only users with more than $500$ followers, in order to focus on successful social capitalists, i.e. ones having effectively gained followers. This means low degree social capitalists have an in-degree between $500$ and $10,000$. The second constraint sets up a minimum of $500$ followees, in order to avoid detecting users whose high overlap index is due only to a very small number of followees.

Dugu\'e \& Perez detected approximately $160000$ social capitalists. Table~\ref{fig:linksanon} shows how they are distributed over the various types of identified behaviors. It is interesting to notice that in this network, most users with more than $10000$ followers are social capitalists ($70\%$). Moreover, users with such a number of followers constitute less than $0.1\%$ of the network. 

\newcolumntype{g}{>{\raggedleft}p{1.5cm}}
\begin{table}[h]
	\centerline{
	\begin{tabular}{ggggr}
			\hline\noalign{\smallskip}
			\textbf{$d^{in}(u)$} & \textbf{$d^{out}(u)$} & \textbf{Number} & \textbf{$r(u)$} & \textbf{$\%$} \\
			\noalign{\smallskip}\hline\noalign{\smallskip}
			\multirow{2}{*}{$> 500$} & \multirow{2}{*}{$>500$} & \multirow{2}{*}{$161424$}  & $> 1$ & $68\%$ \\
			& &   & $[0.7;1]$ & $25\%$ \\ 
			\hline\noalign{\smallskip}
			\multirow{3}{*}{$> 10000$} & \multirow{3}{*}{$>500$} & \multirow{3}{*}{$5743$} & $> 1$ & $66\%$ \\ 
			& &   & $[0.7;1]$ & $25\%$ \\ 
			& &   & $< 0.7$ & $9\%$ \\ 
			\hline\noalign{\smallskip}
		\end{tabular}
	}	
	\caption{Social capitalists detected on the entire Cha \textit{et al}.~\cite{CHBG10} network, with $O(u)>0.74$. \label{fig:linksanon}}
\end{table}

In the experimental part of this work, we decided to use the same method to identify social capitalists in the studied data, instead of the list manually curated by Ghosh \textit{et al}.~\cite{GVK+12}. The reason for this is that the latter is less exhaustive, since it excludes users not following spammers, and does not contain spammers nor bots.  Dugu\'e \& Perez detected approximately $160.000$ social capitalists when Ghosh \textit{et al}.~\cite{GVK+12} detected $100000$ users. Furthermore, some of the listed social capitalists have only a few followers, or only a few reciprocal followers-followee links. Finally, the method proposed by Dugu\'e and Perez~\cite{DP14} detected $80\%$ of the $100000$ social capitalists listed by Ghosh \textit{et al}.

\section{Identifying Community Roles} 
\label{sec:communityroles}
We now present in more details the concept of community role in a complex network. Our work relies on the method proposed by~Guimer\`a \& Amaral~\cite{Guimera2005}.
We first introduce the original measures of Guimer\`a \& Amaral, then highlight their limitations, and finally propose some solutions to these problems.

\subsection{Original approach}
\label{subsec:original}		
In order to characterize the roles of nodes relatively to communities, Guimer\`a \& Amaral~\cite{Guimera2005} defined two complementary measures which allow them to place each node in a 2D role space. Then, they proposed several thresholds to discretize this space, each resulting subspace corresponding to a specific role. In this section, we first describe the measures, then the method used to identify the roles. We then propose a trivial extension to directed networks.

\subsubsection{Role Measures} 
\label{subsubsec:RoleMeasures}		
Both measures are related to the \textit{internal} and \textit{external connectivity} of the node with respect to its community. In other words, they deal with how a node is connected with other nodes inside and outside of its own community, respectively. The first measure, called \emph{within-module degree}, is based on the notion of \emph{$z$-score}. Since the $z$-score will be used again afterwards, we define it here in a generic manner. Let $f(u)$ be any function defined on the vertices, that is $f$ associates a numerical value to any vertex $u$ of the considered graph. The $z$-score $Z_f(u)$ w.r.t. the community of $u$ is defined by:
\begin{eqnarray}
	Z_f(u) & = & \frac{f(u) - \mu_i(f)}{\sigma_i(f)}, u \in C_i	
\label{eq:zscore}		
\end{eqnarray}
\noindent where $C_i$ stands for community number $i$, and $\mu_i(f)$ and $\sigma_i(f)$ respectively denote the mean and standard deviation of $f$ over the nodes belonging to community $C_i$.

Now, let $d_{int}(u)$ be the \textit{internal degree} of a node $u$, i.e. the number of links $u$ has with nodes belonging to its own community. Then, the \textit{within-module degree} of a node $u$, noted $z(u)$ by Guimer\`a \& Amaral~\cite{Guimera2005}, corresponds to the $z$-score of its internal degree. Note that $z$ evaluates the connectivity of a node with its own community, with respect to that of the other nodes of the same community.

The second measure, called \emph{participation coefficient}, is defined as follows:
\begin{eqnarray}
	P(u) & = & 1 - \sum_i \Big ( \frac{d_i(u)}{d(u)} \Big )^2
\label{eq:participation}
\end{eqnarray}
\noindent where $d(u)$ denotes the \emph{total degree} of the node (\emph{i.e.} the number of links it has with any other nodes), and $d_i(u)$ the \textit{community degree} of $u$ (i.e. the number of links it has with nodes of community $C_i$). Note that when $C_i$ corresponds to the community of $u$, then $d_i(u) = d_{int}(u)$. Roughly speaking, the participation coefficient evaluates the connectivity of a node to the community structure in general. If it is close to $0$, then the node is connected to one community only (likely its own). On the contrary, if it is close to $1$, then the node is uniformly linked to a large number of communities.

\subsubsection{Community Roles}
Both measures are used to characterize the \emph{role} of a node within its community. Guimer\`a \& Amaral~\cite{Guimera2005} defined $7$ different roles by discretizing the 2D space formed by $z$ and $P$ using empirically determined thresholds. 

They first used a threshold on the within-module degree, which allowed them to distinguish \emph{hubs} (that is, nodes with $z \geqslant 2.5$) from other nodes, called \textit{non-hubs}. Such hubs are considered as highly linked to their community, when compared to other nodes of the same community. Note that the word hub usually refers to a node with a central position in the whole network, whereas here, the focus is on the community. In other words, in the rest of this article, \textit{hub} implicitly means \textit{community hub}.

\begin{table}[h]
	\centering
	\begin{tabular}{|l|l|l|l|l|}
		\hline
		\multicolumn{4}{|c|}{\textbf{Community role}} & \multicolumn{1}{|c|}{\textbf{External}} \\
		\cline{1-4}
		\multicolumn{2}{|c|}{\textbf{$z$}} & \multicolumn{2}{|c|}{\textbf{$P$}} & \multicolumn{1}{|c|}{\textbf{connectivity}} \\
		\hline
		\multirow{3}{*}{Hub} & \multirow{3}{*}{$\geq 2.5$} & Provincial & $\leq 0.30$ & Low \\
		& & Connector & $ ]0.30;0.75]$ & Strong \\ %
		& & Kinless & $> 0.75$ &  Very strong \\
		\hline
		\multirow{2}{*}{Non-} & \multirow{3}{*}{$< 2.5$} & Ultra-peripheral & $\leq 0.05$ & Very low  \\
		& & Peripheral & $ ]0.05;0.62]$ & Low  \\
		Hub& & Connector & $ ]0.62;0.80]$ &   Strong \\
		& & Kinless & $> 0.80$ &  Very strong \\
		\hline
	\end{tabular}\\[0.4cm] 
	\caption{Guimer\`a \& Amaral's roles and the corresponding $z$ and $P$ thresholds \cite{Guimera2005}.}
	\label{tab:roleDesc}
\end{table}

Those two categories are then subdivided using several thresholds defined on the participation coefficient, as shown in Table~\ref{tab:roleDesc}. By order of increasing $P$, we have: provincial or (ultra-)peripheral, connector and kinless nodes. The two first roles correspond to nodes essentially connected to their community and very (or even completely) isolated from the rest of the network. The third one concerns nodes connected to a number of nodes outside their community. Nodes holding the fourth role are connected to many different communities. Note this is independent from the density of their internal connections: a node can be very well connected in its own community, but not to the rest of the network, in which case it is an ultra-peripheral hub.

\subsubsection{Directed Variants}
\label{sec:directedmeasures}
Many networks representing real-world systems, such as the Twitter follower-followee network we study here, are directed. Of course, it is possible to analyze them through the undirected method, but this would result in a loss of information. 
Yet, extending these measures is quite straightforward: the standard way of proceeding consists in distinguishing incoming and outgoing links. In our case, this results in using $4$ measures instead of $2$: in- and out- versions of both the within-module degree and participation coefficient. 

First, based on the in-degree introduced in section \ref{subsec:measures} and the internal degree from section \ref{subsubsec:RoleMeasures}, let us define the \textit{internal in-degree} of a node, noted $d^{in}_{int}$. It corresponds to the number of incoming links the node has inside its community. By processing the $z$-score of this value, one can derive the \textit{within-module in-degree}, noted $z^{in}$. Similarly, let us note $d^{in}_i$  the \textit{community in-degree}, i.e. the number of incoming links a node has from nodes in community $C_i$. We can now define the \textit{incoming participation coefficient}, noted $P^{in}$, by substituting $d^{in}$ to $d$ and $d^{in}_i$ to $d_i$ in Equation (\ref{eq:participation}). With the same approach, we define $z^{out}$ and $P^{out}$, using the outgoing counterparts $d^{out}$, $d^{out}_{int}$ and $d^{out}_i$. 

In the rest of the article, we call this set of measures the \textit{directed variants}, by opposition to the \textit{original measures} of Guimer\`a \& Amaral~\cite{Guimera2005}.


\subsection{Limitations of this approach} 
We identify two limitations in the approach of Guimer\`a \& Amaral~\cite{Guimera2005}. The first concerns the way the participation coefficient represents the nodes external connectivity, whereas the second is related to the threshold used for the within-module degree.

\subsubsection{External Connectivity}
\label{subsec:participationlimits}
\begin{figure}[h]
	\begin{minipage}{0.32 \linewidth}
		\centerline{\includegraphics[scale=0.10]{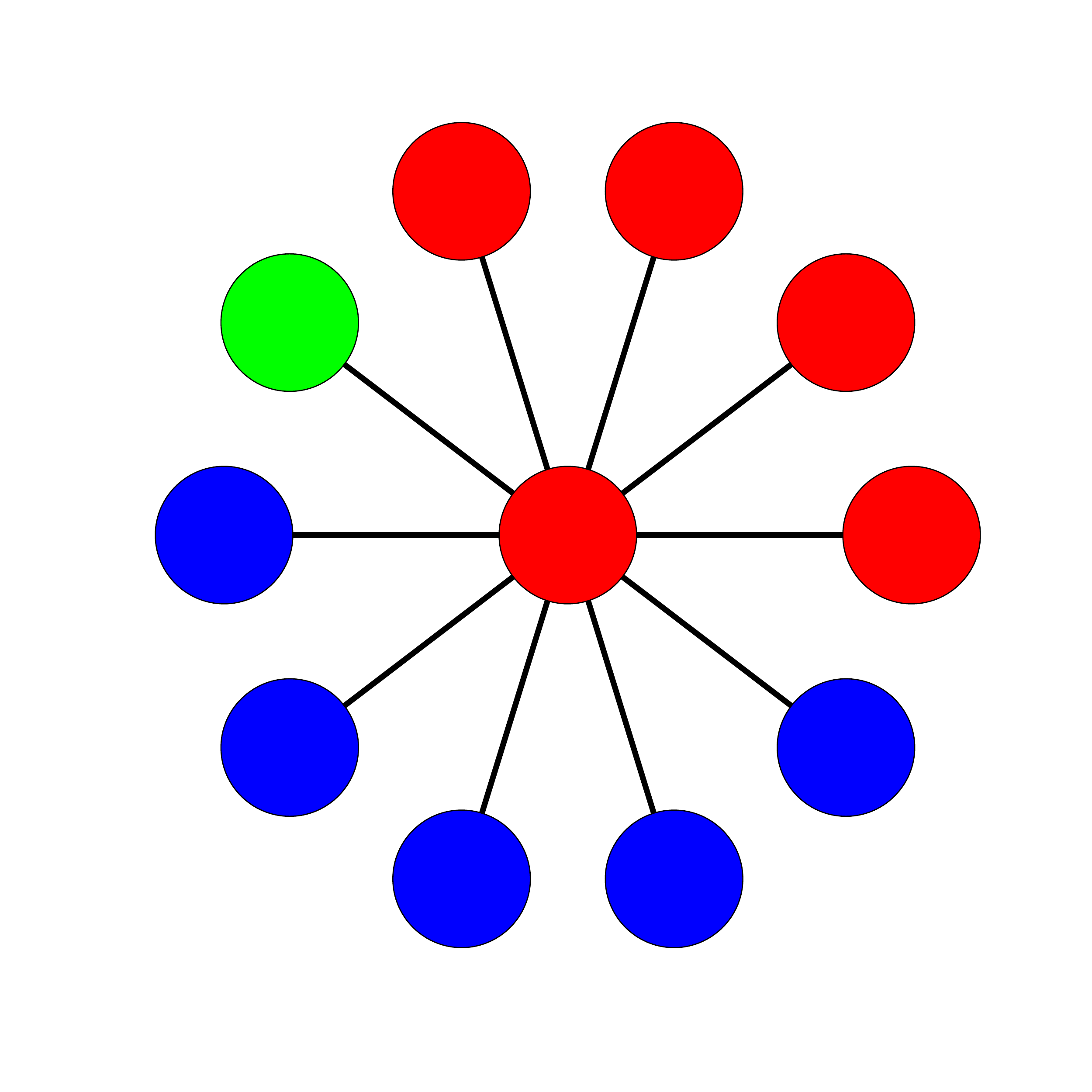}}
	\end{minipage}
	\begin{minipage}{0.32 \linewidth}
		\centerline{\includegraphics[scale=0.10]{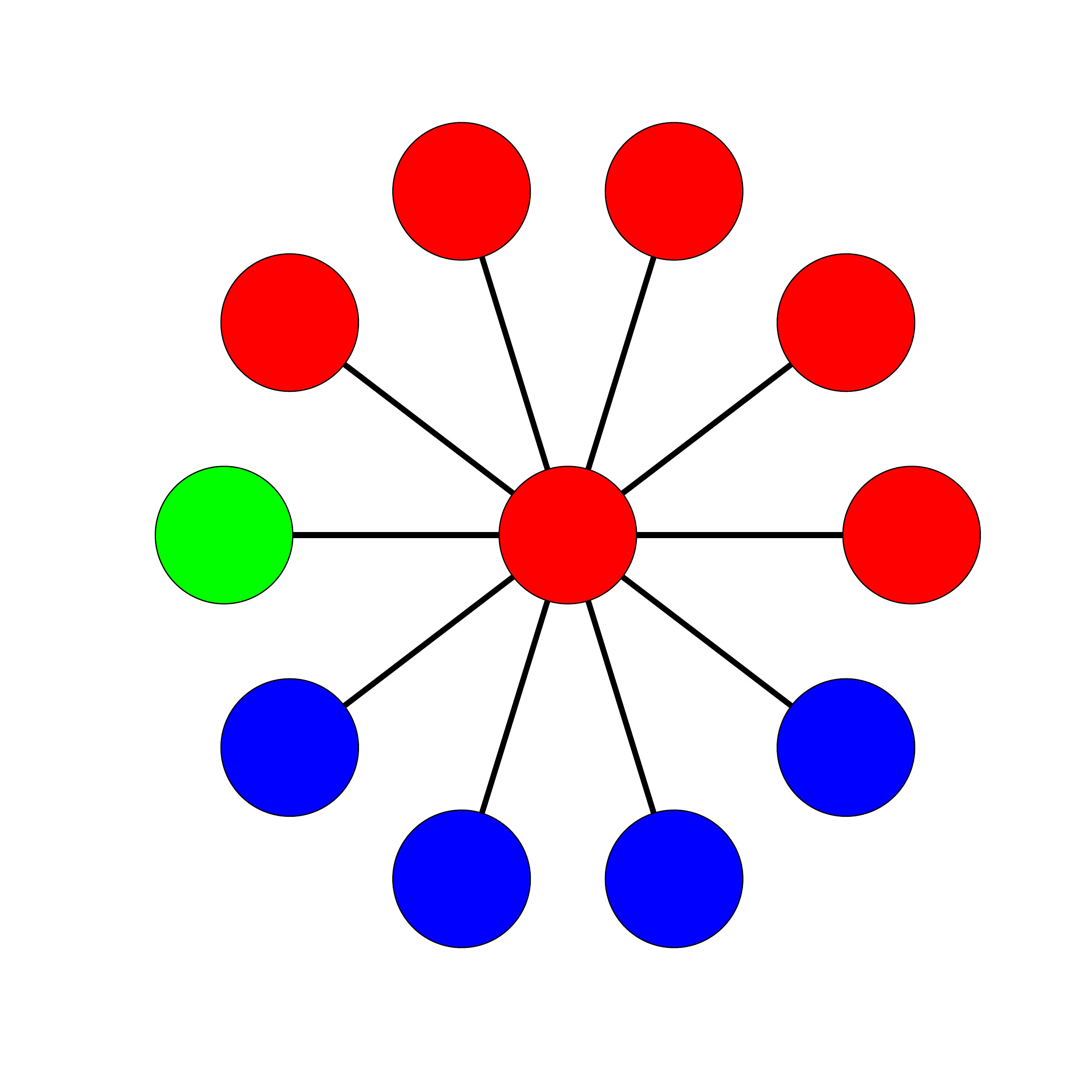}}
	\end{minipage}
	\begin{minipage}{0.32 \linewidth}
		\centerline{\includegraphics[scale=0.10]{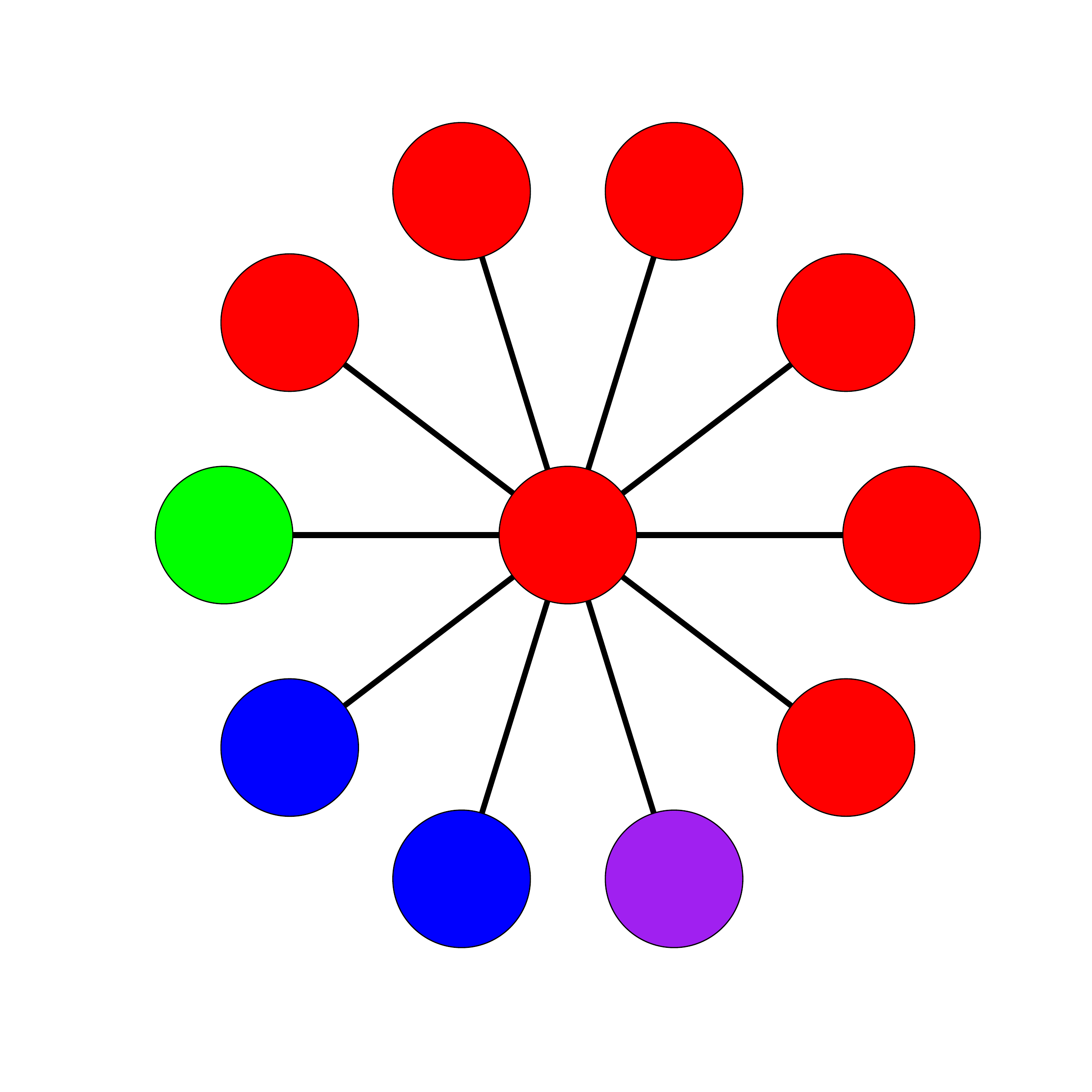}}
	\end{minipage}
	
		\caption{Each color represents a community. In each case, the participation coefficient $P$ of the central node is $0.58$.  \label{fig:participation}}
\end{figure}

We claim that the external connectivity of a given node, i.e. the way it is connected to communities other than its own, can be precisely described in three ways: first, by considering its \textit{diversity}, i.e. the number of concerned communities ; second, in terms of \textit{intensity}, i.e. the number of external links ; and third, relatively to its \textit{heterogeneity}, i.e. the distribution of external links over communities. The participation coefficient combines several of these aspects, mainly focusing on heterogeneity, which lowers its discriminant power. This is illustrated in Figure~\ref{fig:participation}: the external connectivity of the central node is very different in each one of the presented situations. However, $P$ is the same in all cases.

In order to be more illustrative, let us consider two users from our data, which have the same community role according to the original measures. We select two nodes both having a $z$ greater than $2.5$ and a $P$ close to $0.25$. So according to Guimer\`a \& Amaral~\cite{Guimera2005} (see Table~\ref{tab:roleDesc}), they both are provincial hubs, and should have a similar behavior w.r.t. the community structure of the network. However, let us now point out that the first user is connected to $50$ nodes outside its community, whereas the second one has $200000$ connections. This means they actually play different roles in the community structure, either because the second one is connected to much more communities than the first one, or because its number of links with external communities is much larger than for the first user. Similar observations can be made for the directed variants of the participation coefficient. The measures used to define the external connectivity should take this difference into account and assign different roles to these nodes.%

\subsubsection{Fixed Thresholds}
\label{subsec:threshold}
As indicated in the supplementary discussion of~\cite{Guimera2005}, the thresholds originally used to identify the roles were obtained empirically. Guimer\`a \& Amaral first processed $P$ and $z$ for different types of data: metabolic, proteome, transportation, collaboration, computer and random networks. Then, they detected basins of attraction, corresponding to regularities observed over all the studied networks. Each role mentioned earlier corresponds to one of these basins, and the thresholds were obtained by estimating their boundaries. 

Implicitly, these thresholds are supposed to be universal, but this can be criticized. First, Guimer\`a \& Amaral used only one community detection method. A different community detection method can lead to a different community structure, and therefore possibly different basins of attraction. Furthermore, $z$ is not normalized, in the sense it has no fixed boundaries. There is no guarantee the threshold originally defined for this measure will stay meaningful on other networks. As a matter of fact, the values obtained for $z$ in our experiments are far higher for some nodes than the ones observed by Guimer\`a \& Amaral. We also observe that the proportion of nodes considered as hubs (i.e. $z \geq 2.5$) by Guimer\`a \& Amaral is much smaller in our network than in the networks they consider: $0.35\%$ in ours versus $2\%$ in theirs. These thresholds seem to be at least sensitive either to the size of the data, the structure of the network, or to the community detection method.

It is therefore necessary to process new thresholds, more appropriate to the considered data. However, the method used by Guimer\`a \& Amaral~\cite{Guimera2005} itself is difficult to apply, it requires a lot of data. Furthermore, this method assumes that thresholds are universal, which is disproved by our data.

\subsection{Proposed Approach}
\label{sec:proposed}
In this section, we propose some solutions to overcome the limitations of the original approach. First, the participation coefficient mixes several aspects of the external connectivity, which lowers its discriminant power: we introduce several measures to represent these aspects separately. Second, the thresholds used to define the roles do not necessarily hold for all systems: we show how to apply an unsupervised method instead.

\subsubsection{Generalized Measures}
\label{subsec:generalizedmeasures}
In place of the single participation coefficient, we propose $3$ new measures aiming at representing separately the aspects of external connectivity: \emph{diversity}, \emph{intensity} and \emph{heterogeneity}. 
A fourth measure equivalent to the within-degree coefficient is used to describe the internal connectivity.

Because we deal with directed links, each one of these measures exists in two versions: incoming and outgoing (as explained in section \ref{sec:directedmeasures}), effectively resulting in $8$ measures. However, for simplicity matters, we ignore link directions when presenting them in the rest of this section.

All our measures are expressed as $z$-scores (cf. Equation (\ref{eq:zscore})). We know community sizes are generally power-law-distributed, as described in~\cite{Lancichinetti2010}, which means their sizes are heterogeneous. Our community-based $z$-scores allow to normalize the measures relatively to the community size, and therefore to take this heterogeneity into account.


\textbf{Diversity.} The \emph{diversity} $D(u)$ evaluates the number of communities to which a node $u$ is connected (other than its own), w.r.t. the other nodes of its community. This measure does not take into account the number of links $u$ has to each community. Let $\epsilon(u)$ be the number of external communities to which $u$ is connected. The diversity is defined as the $z$-score of $\epsilon$ w.r.t. the community of $u$. It is thus obtained by substituting $\epsilon$ to $f$ in Equation~(\ref{eq:zscore}).

\textbf{External intensity.} The \emph{external intensity} $I_{ext}(u)$ of a node $u$ measures the amount of links $u$ has with communities other than its own, w.r.t. the other nodes of its community. Let $d_{ext}(u)$ be the external degree of $u$, that is the number of links $u$ has with nodes belonging to another community than its own. The external intensity is defined as the $z$-score of the external degree, i.e. we obtain it by substituting $d_{ext}$ to $f$ in Equation~(\ref{eq:zscore}). 

\textbf{Heterogeneity.} The \emph{heterogeneity} $H(u)$ of a node $u$ measures the variation of the number of links a node $u$ has, from one community to another. To that aim, we compute the standard deviation of the number of links $u$ has to each community. We note this value $\delta(u)$. The heterogeneity is thus the $z$-score of $\delta$ w.r.t. the community of $u$. As previously, it can be obtained by substituting $\delta$ to $f$ in Equation~(\ref{eq:zscore}).

\textbf{Internal intensity.} In order to represent the internal connectivity of the node $u$, we use the $z$ measure of Guimer\`a \& Amaral~\cite{Guimera2005}. Indeed, it is based on the notion of $z$-score, and is thus consistent with our other measures. Moreover, we do not need to add measures such as diversity or heterogeneity, since we consider one node can belong only to one community. Due to the symmetry of this measure with the external intensity, we refer to $z$ as the \emph{internal intensity}, and note it $I_{int}(u)$. 

\subsubsection{Unsupervised Role Identification}
\label{sec:unsupervised}		
Our second modification concerns the way roles are defined. As mentioned before, the thresholds defined by Guimer\`a \& Amaral~\cite{Guimera2005} are not necessarily valid for all data. Moreover, the consideration of link directions and our generalization of the measures invalidate the existing thresholds, since we now have $8$ distinct measures, all different from the original ones. We could try estimating more appropriate thresholds, but as explained in section \ref{subsec:threshold}, the method originally used by Guimer\`a \& Amaral~\cite{Guimera2005} to estimate their thresholds is impractical since it requires a certain amount of data. The fact our measures are all $z$-scores also weakens the possibility to get thresholds applicable to all systems, which means the estimation process should potentially be performed again for each studied system.

To overcome these problems, we propose to apply an automatic method instead, by using unsupervised classification. First, we process all the measures for the considered data. Then, a cluster analysis method is applied. Each one of the clusters identified in the $8$-dimensional role space is considered as a community role. This method is not affected by the number of measures used, and allows to adjust thresholds to the studied system. If the number of roles is known in advance, for instance because of some properties of the studied system, then one can use an appropriate clustering method such as $k$-means, which allows specifying the number $k$ of clusters to find. Otherwise, it is possible to use cluster quality measures to determine which $k$ is the most appropriate ; or to apply directly a method able to estimate at the same time the optimal number of clusters and the clusters themselves.

\section{Community Roles of Social Capitalists}
\label{sec:results}		
In this section, we present the results we obtained on a Twitter network using the methods presented in Section \ref{sec:communityroles}. We first introduce the data and tools we used, then the roles we identified. We then focus social capitalists and the roles they hold.

\subsection{Data and Tools}
We analyze a freely-available anonymized Twitter follower-followee network, collected in 2009 by Cha \textit{et al}.~\cite{CHBG10}. It contains about $55$ million nodes representing Twitter users, and almost $2$ billion directed links corresponding to follower-followee relationships. We had to consider the size of these data when choosing our analysis tools. 

For community detection, we selected the Louvain method \cite{Blondel2008}, because it is widespread and proved to be very efficient when dealing with large networks. We retrieved the C++ source code published by its authors, and adapted it in order to optimize the directed version of the modularity measure, as defined by Leicht and Newman~\cite{Newman2008}. Empirical benchmarks show that our adapted version performs better than the original one on directed network. All the role measures, that is Guimer\`a \& Amaral's original measures, their directed variants (section \ref{subsec:original}) and our new measures (section \ref{sec:proposed}), were computed using the community structure detected through these means. We also implemented them in C++, using the same sparse matrix data structure than the one used in the Louvain method. 

All resulting values were normalized, in order to avoid scale difference problems when conducting the cluster analysis. The clustering was performed using an open source implementation of a distributed version of $k$-means \cite{Liao2009}. 
Since we do not know the expected number of roles, we applied this algorithm for $k$ ranging from $2$ to $15$, and selected the best partition in terms of Davies-Bouldin index~\cite{Davies1979}. We selected this index because it is a good compromise between the reliability of the estimated quality of the clusters, and the computing time it requires. All pre- and post-processing scripts related to the cluster analysis were implemented in R.
The whole source code is freely available online\footnote{  
\url{https://github.com/CompNet/Orleans}}. 
Because we are dealing with $55$ millions of objects and only $8$ attributes, we know that a lot of local maximums exist while minimizing the within-cluster sum of squares with the $k$-means algorithm. In this paper, we are specifically looking for assessing the social capitalists visibility in the Twitter network. To achieve this goal, we do not necessarily need to get the best partition of the role space. We are actually looking for a partition in which clusters are well separated and interpretable according to Guimer\`a \& Amaral's terminology. With such a partition, it is possible to look at the specific roles held by social capitalists in these clusters and to determine their actual visibility. 

\subsection{Roles Expected for Social Capitalists} 
\label{subsec:expectedrole}	

We expect the degree of social capitalists to play an important role considering their position (see Section~\ref{sec:cap}). High in-degree social capitalists (namely greater than $10000$) should be well connected to their communities -hubs- or to the other communities -connectors, or both. Being connectors would indicate they obtained a high visibility on the whole network and not only in their own communities.

Furthermore, because we take the direction of links into account in our measures, we expect social capitalists to be discriminated according to their ratio, i.e. the number of outgoing links divided by the number of incoming links. We especially expect high in-degree social capitalists with a small ratio (so-called \emph{passive social capitalists} according to~\cite{DP14}) to be highly connected to their communities and to the rest of the graph.
Considering low degree social capitalists, it is not possible to predict their roles without any further information. The study will thus be of great interest to characterize their visibility.



\subsection{Detected Roles}
For the sake of completeness, we first used the directed measures (section \ref{subsec:original}) of Guimer\`a \& Amaral~\cite{Guimera2005}. As mentioned in Section \ref{subsec:threshold}, the threshold they defined for $z$ is irrelevant for our data. Furthermore, this threshold was determined for an undirected version. So we adopted here the unsupervised role identification method we proposed (section~\ref{sec:unsupervised}). 

\subsubsection{Directed Variants}
A correlation study shows $z^{out}$ and $z^{in}$ are slightly correlated (with a correlation coefficient $\rho<0.3$), whereas the correlation is zero for all other pairs of measures. This seems to confirm the interest of considering link directions in the role measures. When doing the cluster analysis, the most separated clusters are obtained for $k = 6$. 
An \textsc{ANOVA} followed by \textit{post hoc} tests ($t$-test with Bonferroni's correction) showed significant differences exist between all clusters and for all measures. 
%

An analysis of the distribution of high in-degree social capitalists in these clusters shows that a few of these users occupy a connector hub role. This is quite expected as said in section~\ref{subsec:expectedrole}. However, most of the high in-degree social capitalists are considered as non-hubs and peripheral or ultra-peripheral nodes. More than $60\%$ of the users with a high ratio are classified as ultra-peripheral nodes for both incoming and outgoing directions, which is rather surprising since they have a really high degree. However, they are classified in a cluster with low $z$ and $P$ (both in- and out- versions). The low $z$ indicates these users are not much connected to their community (relatively to the other nodes of the same community), and must thus be more connected to other communities. Still, $P$ does not highlight this aspect of their community-related connectivity, and they appear as peripheral. This inconsistency of the detected roles confirms the limitations of $P$ described in section \ref{subsec:participationlimits}. 

\subsubsection{Generalized Measures}
The correlation between the generalized measures is very low overall, ranging from almost $0$ to $0.4$. In particular, both versions of the same measure (incoming vs. outgoing) are only slightly correlated, which is another confirmation of the interest of considering link directions. Only three measures are strongly correlated: internal and external intensities and heterogeneity ($\rho$ ranging from $0.78$ to $0.92$). The relation between both intensities seems to indicate that variations on the total degree globally affect similarly internal and external degrees. The very strong correlation observed between heterogeneity and intensity means only nodes with low intensity are homogeneously connected to external communities, whereas nodes with many links are connected heterogeneously.

Similarly to the directed measures, the most separated clusters are obtained with $k = 6$. These $6$ clusters are given in Table~\ref{tab:groupes_generalized} with their sizes and roles. However, the correspondance with the original nomenclature is rougher, since these measures are farther from the original ones. The average of each measure per cluster is showed in Table~\ref{tab:moyennes_generalized}. Like before, \textsc{ANOVA} and \textit{post hoc} tests showed significant differences between all clusters and for all measures. We now conduct a detailed analysis of the different roles we obtain. 

\begin{table}[h]
	\centering
	\begin{tabular}{|c|r|r|r|}
		\hline
		\textbf{C} & \textbf{Size} & \textbf{Proportion} & \textbf{Role} \\
		\hline
		1 & $24543667$ & $46.68\%$ & Ultra-peripheral non-hubs \\
		2 &      $304$ & $<0.01\%$ & Kinless hubs \\
		3 &   $303674$ &  $0.58\%$ & Connector hubs	\\
		4 & $11929722$ & $22.69\%$ & Incoming Peripheral non-hubs \\
		5 & $10828599$ & $20.59\%$ & Outgoing Peripheral non-hubs \\
		6 &  $4973717$ &  $9.46\%$ & Connector non-hubs \\
		\hline
	\end{tabular}\\[0.4cm]
	\caption{Clusters detected with the generalized measures: cluster number $C$ used in the paper, sizes in terms of node count and proportion of the whole network, and roles according to the Guimer\`a \& Amaral~\cite{Guimera2005} nomenclature.}
	\label{tab:groupes_generalized}
\end{table}

\noindent \textbf{Cluster 1.} Because both internal intensity versions (equivalent to $z$) are negative, nodes in this cluster cannot be hubs. The negative external measures indicate these nodes are not connectors either. We can thus consider them as ultra-peripheral non-hubs. This cluster is the largest one, with $47\%$ of the network nodes. This confirms the matching with this role, whose nodes constitute generally most of the network.

\noindent \textbf{Clusters 4 and 5.} Cluster $4$ is very similar to Cluster $1$. However, its incoming diversity is $0.69$. These nodes are again peripheral, because the external intensity is negative. Still, incoming links come from a larger number of communities.
Cluster $5$ is also similar to Cluster $1$. However, both versions of diversity are positive for this cluster, with an outgoing diversity of $0.60$. External links are thus connected to a larger number of communities.
Clusters $4$ and $5$ are the second ($23\%$) and third ($21\%$) largest ones, respectively. By gathering all the peripheral and ultra-peripheral nodes, we obtain $91\%$ nodes of the network.

\noindent \textbf{Cluster 6.} The internal intensity is still close to $0$ but positive. Thus, these nodes are non-hubs, even if they are more connected to their community than those of the previous clusters. Like the other external measures, the external intensity is low but still positive. These nodes are relatively well-connected to other communities, and we can therefore consider them as connectors. Both versions of the diversity are relatively high, which indicates these nodes are not only more connected to their community as well as others, but also to a larger number of distinct communities.

\noindent \textbf{Cluster 3.} The high internal intensity allows us to state that these nodes are hubs. Furthermore, the high external measures indicate these nodes are connected to a high number of nodes from a lot of other communities, and thus are connector hubs. Notice outgoing measures are higher. This cluster represents only $0.6\%$ of the network, meaning this role is very uncommon. 

\noindent \textbf{Cluster 2.} This observation is even more valid for Cluster $2$, which represents much less than $1\%$ of the nodes. For this cluster, all measures are really high. The incoming versions are always higher than their outgoing counterparts. We call these users kinless hubs according to Guimer\`a \& Amaral's nomenclature.

\begin{table}[h]
	\centering
	\begin{tabular}{|c|r|r|r|r|}
		\hline
		$\mathbf{Cluster}$ & $\mathbf{I^{out}_{int}}$ & $\mathbf{I^{in}_{int}}$ & $\mathbf{D^{out}}$ & $\mathbf{D^{in}}$\\
		\hline
		1 & $-0.12$ &  $-0.03$ & $-0.55$ & $-0.80$ 	\\
		2 & $94.22$ & $311.27$ &  $7.18$ & $88.40$ 	\\
		3 &  $5.52$ &   $1.40$ &  $5.60$ &  $3.10$ 	\\
		4 & $-0.04$ &   $0.00$ & $-0.37$ &  $0.69$ 	\\
		5 & $-0.03$ &  $-0.01$ &  $0.60$ &  $0.19$ 	\\
		6 &  $0.48$ &   $0.12$ &  $1.96$ &  $1.70$ 	\\
		\hline
	\end{tabular}\\[0.2cm]
	\begin{tabular}{|c|r|r|r|r|}
		\hline
		$\mathbf{Cluster}$ &
 		$\mathbf{I^{out}_{ext}}$ & $\mathbf{I^{in}_{ext}}$ & $\mathbf{H^{out}}$ & $\mathbf{H^{in}}$ \\
		\hline
		1 & $-0.09$ &  $-0.04$ &  $-0.12$ &  $-0.06$ 	\\
		2 & $113.87$ & $283.79$ & $112.79$ & $285.57$ 	\\
		3 &  $5.28$ &   $1.43$ &   $6.76$ &   $2.34$ 	\\
		4 & $-0.07$ &   $0.00$ &  $-0.10$ &  $-0.01$ 	\\
		5 & $-0.03$ &  $-0.02$ &  $-0.04$ &  $-0.02$ 	\\
		6 & $0.35$ &   $0.12$ &   $0.53$ &   $0.19$ 	\\
		\hline
	\end{tabular}\\[0.4cm]
	
	\caption{Average generalized measures obtained for the $6$ detected clusters 
	\label{tab:moyennes_generalized}.}
\end{table}

It is worth noticing that, whatever the considered measures, some of the roles defined by Guimer\`a \& Amaral~\cite{Guimera2005} are not represented in the studied network. This is consistent with the remarks previously made for other data by Guimer\`a \& Amaral~\cite{Guimera2005}, and  confirms the necessity of having an unsupervised approach to define roles in function of measures. It is also consistent with the strong correlation observed between internal and external intensities: missing roles would be nodes possessing a high internal intensity but a low external one, or vice-versa. However, those are very infrequent in our network.

\subsection{Relations between clusters}
\label{subsec:relations}
\begin{figure}
	\center	
	\includegraphics[scale=0.54]{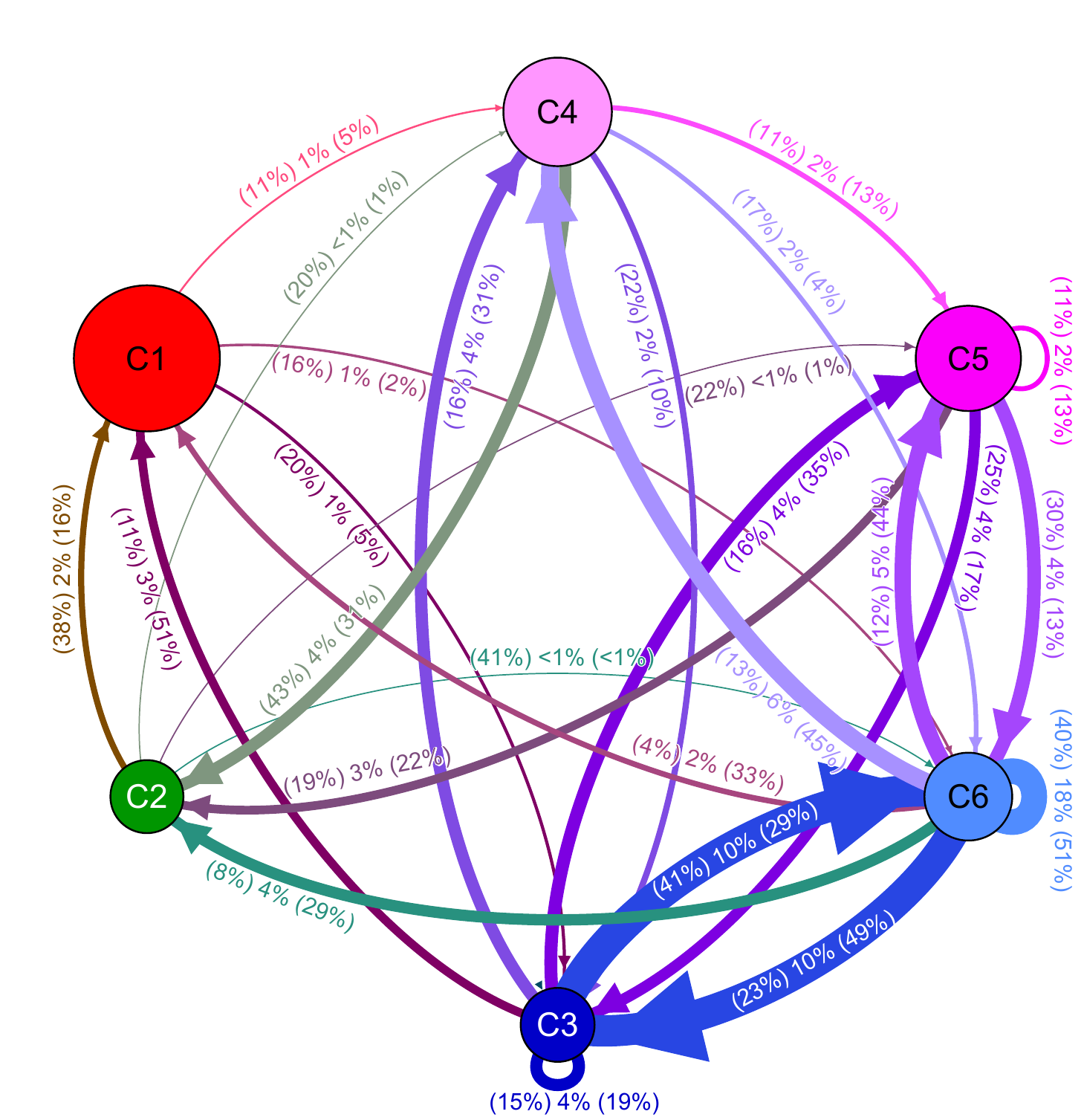}
	\caption{Interconnection between clusters. A vertex $C_i$ corresponds to Cluster $i$ from Table \ref{tab:groupes_generalized}. An arc $(i,j)$ represents the set of links connecting nodes from Cluster $i$ to nodes from Cluster $j$, labeled with $3$ values. Each value describes which proportion of links the arc represents, relatively to 3 distinct sets: first relatively to all links starting from Cluster $i$, second relatively to all links in the whole network, and third relatively to all links ending in Cluster $j$. The arc thickness is proportional to the second value, the vertex size to the number of nodes the corresponding cluster contains. For matters of readability, arcs representing less than $1\%$ of the network links and $10\%$ of the cluster links are not displayed.}\label{fig:ctoc}
\end{figure}

We now discuss how the nodes are connected depending on the role they hold. Figure~\ref{fig:ctoc} is a simplified representation of this interconnection pattern. 

The outgoing links of ultra-peripheral (Cluster 1) and peripheral (Clusters 4 and 5) nodes target mainly kinless hubs (Cluster 2) and connectors (Clusters 3 and 6), representing $74\%$ (Cluster 1), $82\%$ (Cluster 4), and $74\%$ (Cluster 5) of their connections. These (ultra-)peripheral nodes, which are the most frequent in the network, thus mainly follow very connected users, probably the most influent and relevant ones. This seems consistant: they follow only a few users, and so choose the most visible ones. 

Connector nodes (Clusters 3 and 6) are mainly linked to other connectors nodes. They have the tightest connection, since their arcs amounts to a total of $43\%$ of the network links. This is worth noticing, because these clusters are far from being the largest ones. They are also largely connected to the rest of the clusters too, especially with outgoing links. 
Connectors follow massively users of all clusters, so we suppose they constitute the backbone of the network.

Kinless hubs (Cluster 2) are massively followed by non-hubs, representing $38\%$ (Cluster 1), $43\%$ (Cluster 4), $19\%$ (Cluster 5) and $8\%$ (Cluster 6) of these Clusters' outgoing links. And interestingly, the links coming from kinless hubs target the same clusters: $9\%$ go to Cluster 1, $20\%$ to Cluster 4, $22\%$ to Cluster 5 and $41\%$ to Cluster 6. This means the most visible and popular nodes of the network mostly follow and are followed by much less popular users. One could have expected the network to be hierarchically organized around roles, with more peripheral nodes connected to less peripheral nodes. But this is clearly not the case. First, (ultra-)peripheral nodes are marginally connected to other nodes holding the same role, they prefer to follow connectors and/or hubs. Second, kinless and connector hubs, although well connected to connector non-hubs, do not have direct links, i.e. these users do not follow each other.

\subsection{Position of Social Capitalists}
\label{sec:position}
As stated previously, we use a list of approximately $160000$ social capitalists detected by Dugu\'e \& Perez~\cite{DP14}. In the following, we analyze how social capitalists are distributed am\-ongst the detected roles. As explained in Section~\ref{sec:cap}, we split social capitalists according to their in-degree (number of followers). Recall that \emph{low in-degree social capitalists} have an in-degree between $500$ and $10000$, and \emph{high in-degree social capitalists} an in-degree greater than $10000$. These social capitalists are known for having especially well succeeded in their goal of gaining visibility.

The tables in this section describe how the various types of social capitalists are distributed over the clusters.
In each cell, the first row is the proportion of social capitalists belonging to the corresponding cluster, and the second one is the proportion of cluster nodes which are social capitalists. Values of interest are indicated in bold and discussed in the text.

\subsubsection{Low in-degree social capitalists}
Low in-degree social capitalists are mostly assigned to three clusters: $3$, $5$ and $6$ (see Table~\ref{tab:ksociaux500_generalized}). Most of them belong to Cluster $6$, which contains non-hub connector nodes. These nodes, which have only slightly more external connections than the others, are nevertheless connected to far more communities. Social capitalists in this cluster seem to have applied a specific strategy consisting in creating links with many communities. This strategy is still not completely working, though, as shown by the relatively low external incoming intensity (meaning they do not have that many followers). 

Nodes from Cluster $3$ are connector hubs, who follow more users than the others. Because \textbf{IFYFM} social capitalists have a ratio greater than $1$ and thus more followees than followers, it is quite intuitive to observe that they are twice as many than the other users in this cluster. The high outgoing diversity of Cluster $3$ tells us that these social capitalists follow users from a large variety of communities, not only theirs (to which they are well connected). The high external outgoing intensity shows that these users massively engage in the \textbf{IFYFM} process, but did not yet receive a lot of following back, as shown by their low external incoming intensity.
Finally, roughly $20\%$ of social capitalists with ratio $r < 1$ belong to Cluster $5$, which contains non-hub peripheral nodes. This shows that a non-neglictible share of social capitalists are isolated relatively to both their community and the other ones. 

\begin{table}[h]
	\centering
	\begin{tabular}{|l|r|r|r|r|r|r|}
		\hline
		\textbf{Ratio} & \textbf{Cluster 1}  & \textbf{Cluster 2} & \textbf{Cluster 3} \\
		\hline
	  	\multirow{2}{*}{$r \leq 1$} & $0.01\%$ & $0.00\%$ & $\mathbf{23.10\%}$   \\
		& $< 0.01\%$ & $0.00\%$ & $3.71\%$    \\
		\hline
		\multirow{2}{*}{$r > 1$}  & $0.03\%$   & $0.00\%$ & $\mathbf{18.78\%}$   \\
		& $< 0.01\%$ & $0.00\%$ & $\mathbf{6.61}\%$  \\
		\hline
	\end{tabular}\\[0.2cm]
	\begin{tabular}{|l|r|r|r|r|r|r|}
		\hline
		\textbf{Ratio}  & \textbf{Cluster 4} & \textbf{Cluster 5} & \textbf{Cluster 6}\\
		\hline
		\multirow{2}{*}{$r \leq 1$}   &  $3.42\%$ &  $\mathbf{18.28\%}$ & $\mathbf{55.19\%}$  \\
		&  $0.14\%$ &  $0.08\%$ & $0.54\%$  \\
		\hline
		\multirow{2}{*}{$r > 1$}    &  $0.48\%$   &  $\mathbf{14.31\%}$ & $\mathbf{66.40\%}$ \\
		& $< 0.01\%$ &   $0.14\%$ 				& $1.43\%$ \\
		\hline
	\end{tabular}\\[0.4cm]
	\caption{Distribution of low in-degree social capitalists over clusters obtained from the generalized measures.
	\label{tab:ksociaux500_generalized}}
\end{table}

These observations show that most of these users are deeply engaged in a process of soliciting users from other communities, not only theirs. Some of them are even massively following users from a wide diversity of communities. This tends to show that these users may obtain an actual visibility across many communities of the network by spreading their links efficiently.

\subsubsection{High in-degree social capitalists}
Most of the high in-degree social capitalists are gathered in Cluster $3$ (see Table~\ref{tab:ksociaux10000_generalized}), corresponding to connector hubs. This is consistent with the fact these users have a high degree. Users of Cluster $3$ have a high outgoing diversity and a high outgoing external intensity: this shows they practice the \textbf{IFYFM} strategy actively, by following a lot of users from a wide range of communities.
The rest of these users is contained in Cluster $2$. Nodes in these clusters are kinless hubs and can thus be considered as successful users. Indeed, they are massively followed by a very high number of users from an extremely large variety of communities.
Only high in-degree social capitalists with a ratio smaller than $0.7$ and a few with a ratio smaller than $1$ are classified in this cluster. This is consistent with the roles one could expect for social capitalists (Section~\ref{subsec:expectedrole}).

\begin{table}[h]
	\centering
	\begin{tabular}{|l|r|r|r|}
		\hline
		\textbf{Ratio} & \textbf{Cluster 1}  & \textbf{Cluster 2} & \textbf{Cluster 3} \\
		\hline
		\multirow{2}{*}{$r \leq 0.7$} & $0.00\%$ & $\mathbf{12.14\%}$ & $\mathbf{87.29\%}$ \\
		& $0.00\%$ & $\mathbf{21.05\%}$ & $0.15\%$ \\
		\hline
		\multirow{2}{*}{$0.7 < r \leq 1$} & $0.00\%$ & $1.55\%$ & $\mathbf{95.64\%}$ \\
		& $0.00\%$ & $\mathbf{7.24\%}$ & $0.45\%$ \\
		\hline
		\multirow{2}{*}{$r > 1$} & $0.00\%$ & $0.03\%$ & $\mathbf{97.99\%}$ \\
		& $0.00\%$ & $0.33\%$ & $1.22\%$ \\
		\hline
	\end{tabular}\\[0.2cm]
	\begin{tabular}{|l|r|r|r|}
		\hline
		\textbf{Ratio} & \textbf{Cluster 4} & \textbf{Cluster 5} & \textbf{Cluster 6}\\
		\hline
		\multirow{2}{*}{$r \leq 0.7$}  & $0.00\%$  & $0.00\%$ & $0.57\%$ \\
		&  $0.00\%$ & $0.00\%$ & $< 0.01\%$ \\
		\hline
		\multirow{2}{*}{$0.7 < r \leq 1$}  & $0.00\%$ & $0.00\%$ & $2.81\%$ \\
		& $0.00\%$ &  $0.00\%$ & $< 0.01\%$ \\
		\hline
		\multirow{2}{*}{$r > 1$}  & $0.00\%$ & $0.00\%$ & $1.98\%$ \\
		& $0.00\%$ & $0.00\%$ & $< 0.01\%$ \\
		\hline
	\end{tabular}\\[0.4cm]
	\caption{Distribution of high in-degree social capitalists over clusters obtained from the generalized measures.
	\label{tab:ksociaux10000_generalized}}
\end{table}
These observations mean that most of these users are well connected in their communities but also with the rest of the network. This shows the efficiency of these users strategies. Indeed, most of the users are linked to a wide range of communities, and thus reach a high visibility in a large part of the network.

\section{Related Works}
\label{sec:related}


The notion of role in network science first appeared in the seventies. Two nodes are considered as holding the same roles if they are structurally equivalent~\cite{Lorrain1971}, namely if they share the same neighbors in the graph representing their relations~\cite{Burt1990}. This notion also appears in block models, where networks are partitioned as groups sharing the same patterns of relations~\cite{Holland1983}. In both cases, the concept of role is defined globally, i.e. relatively to the whole network.

More recently, Guimer\`a \& Amaral introduced the concept of \textit{community} role to study metabolic networks \cite{Guimera2005}, by considering node connectivity at the level of the community structure, i.e. an intermediate level. As explained in Section \ref{subsec:original}, they fist apply a standard community detection method, and then characterize each node according to two \textit{ad hoc} measures, each one describing a specific aspect of the community-related connectivity. The first expresses the intensity of its connections to the rest of its own community, whereas the second quantifies how uniformly it is connected to all communities. 
The node role is then selected among $7$ predefined ones by comparing the two values to some empirically fixed thresholds. Guimer\`a \& Amaral showed certain systems possess a role invariance property: when several instances of the system are considered, nodes are different but roles are similarly distributed. 

Scripps \textit{et al}.~\cite{Scripps2007}, apparently unaware of Guimer\`a \& Amaral's work, later adopted a similar approach, but this time for influence maximization and link-based classification purposes. They also use two measures: first the degree, to assess the intensity of the general node connectivity, and second an \textit{ad hoc} measure, to reflect the number of communities to which it is connected. They then use arbitrary thresholds to define $4$ distinct roles. 

Even more recently, Klimm \textit{et al}.~\cite{Klimm2014} criticized Guimer\`a \& Amaral's approach, and proposed a modification based on two different measures. They first defined the \textit{hubness index}, which compares the degree of a node $u$ with the probability for this node to have the same number of links in a subgraph with fixed density and size. Their \textit{local hubness index} is a variant using the density and size of the community containing $u$ while their \textit{global hubness index} uses the whole network density and size. They claim normalizing the internal degree with this method (using density and size) leads to better results than with the $z$-score used by Guimer\`a \& Amaral. However, the expected improvement is not clearly shown in the article. The second measure is a modification of the participation coefficient, taking the form of a normalized vector representing the participation of a node to each community of the network. They also introduced a \textit{dispersion index} that is a normalized vector representing the participation of a node to each community he is connected to. The limitations we highlighted for the original participation coefficient are also valid for these two variants: none of them is able to model all aspects of the external connectivity of a node. The first measure still encapsulates all aspects of the external connectivity, while the second one deals simultaneously with its heterogeneity and diversity (cf. Section \ref{subsec:participationlimits}). Furthermore, Klimm \textit{et al.} do not propose a method to assign roles to nodes according to their measures, even empirically, they only analyze a few small biological networks.

\section{Conclusion}
\label{sec:conclusion}		
In this article, our goal is to characterize the position of social capitalists in Twitter. For this purpose, we propose an extension of the method defined by Guimer\`a \& Amaral~\cite{Guimera2005} to characterize the community role of nodes in complex networks. We first define directed variants of the original measures, and extend them further in order to take into account the different aspects of node connectivity. Then, we propose an unsupervised method to determine roles based on these measures. It has the advantage of being independant from the studied system. Finally, we apply our tools to a follower-followee Twitter network. We find out the different kinds of social capitalists occupy very specific roles. Those of low in-degree are mostly connectors non-hubs. This shows they are engaged in a process of spreading links across the whole network, and not only their own community. Those of high in-degree are classified as kinless or connectors hubs, depending on their ratio $r$. This shows the efficiency of their strategies, which lead to a high visibility for a vast part of the network, not only for their own community.

The most direct perspective for our work is to assess its robustness. In particular, it is important to know how the stability of the detected communities and clusters affects the identified roles. In this study, our aim was to assess the social capitalist visibility, which is independant of this goal. Furthermore, the very large size of the data prevented us to do so efficiently. On a related note, we want to apply our method to other smaller systems, in order to check for its general relevance. The method itself can also be extended in two ways. First, it would be relatively straightforward to take link weights into account (although this was not needed for this work). Second, and more interestingly, it is also possible to adapt it to overlapping communities (by opposition to the mutually exclusive communities considered in this work) in a very natural way, by introducing additional internal measures symmetrical to the existing external ones. This could be a very useful modification when studying social networks, since those are supposed to possess this kind of community structures, in which a node can belong to several communities at once \cite{Arora2012}.

\bibliographystyle{plain}
\bibliography{biblio}

\end{document}